\RequirePackage[l2tabu, orthodox]{nag}
\RequirePackage{snapshot}

\documentclass[9pt,onecolumn]{extarticle}

\makeatletter
\if@twocolumn
  \usepackage[dvips,letterpaper,top=0.5in, bottom=0.5in, left=0.75in, right=0.5in,includefoot,heightrounded]{geometry}
\else
  \usepackage[dvips,letterpaper,margin=1in,includefoot,heightrounded]{geometry}
\fi

\usepackage{srcltx}

\usepackage[russian,portuges,english]{babel}

\iflanguage{portuges}
    {\newcommand{\keywordname}{Palavras-chaves}}
    {\newcommand{\keywordname}{Keywords}}

\usepackage{amsmath}
\usepackage{amssymb,amsfonts}

\usepackage{abstract}

\usepackage{graphicx}
\usepackage[usenames,dvipsnames,svgnames,x11names]{xcolor}
\usepackage{subfigure}

\usepackage{booktabs}

\usepackage{setspace}
\usepackage{flushend}

\usepackage{cite}

\usepackage{hyperref}
\usepackage[normalem]{ulem}

\usepackage{enumerate}

\usepackage{multirow}
\usepackage[noend]{algpseudocode}

\usepackage{listings}

\usepackage[short,12hr]{datetime}
       \usepackage{fouriernc}
\makeatletter

\newcommand{\printtitle}{%
\makeatletter
\if@twocolumn

\twocolumn[%
  \maketitle
  \begin{onecolabstract}
    \myabstract
  \end{onecolabstract}
  \begin{center}
    \small
    \textbf{\keywordname}
    \\\medskip
    \mykeywords
  \end{center}
  \bigskip
]
\saythanks
\else
  \maketitle
  \begin{onecolabstract}
    \myabstract
  \end{onecolabstract}
  \begin{center}
    \small
    \textbf{\keywordname}
    \\\medskip
    \mykeywords
  \end{center}
  \bigskip
  \onehalfspacing
\fi
\makeatother
}

\author{%
B.~G.~Palm%
\thanks{Programa de P\'os-gradua\c{c}\~ao em Estat\'istica,
Universidade Federal Pernambuco, Brazil;
\protect\url{brunagpalm@gmail.com}
}
\and
D.~I.~Alves%
\thanks{Federal University of PAMPA, Brazil
and
	Federal University of Santa Catarina, Brazil
\protect\url{dimasalves@unipampa.edu.br }
}
\and
M.~I.~Pettersson%
\thanks{Department of
	Mathematics
	and Natural Sciences,
	Blekinge Institute of Technology,
	Sweden;
	\protect\url{mats.pettersson,viet.thuy.vu@bth.se}
}
\and
V.~T.~Vu${}^\ddagger$%
\and
R.~Machado%
\thanks{Department of
	Telecommunications,
	Aeronautics Institute
	of Technology (ITA),
	S\~ao Jos\'e dos Campos - SP,
	Brazil;
	\protect\url{rmachado@ita.br}
}
\and
R.~J.~Cintra%
\thanks{Signal Processing Group,
Departamento de Estat\'istica,
Universidade Federal Pernambuco, Brazil;
\protect\url{rjdsc@de.ufpe.br}
}
\and
F.~M.~Bayer%
\thanks{Departamento de Estat\'{\i}stica
and LACESM,
Universidade Federal de Santa Maria, Brazil;
\protect\url{bayer@ufsm.br}
}
\and
P.~Dammert%
\thanks{Saab Electronic Defence Systems, Sweden;
\protect\url{patrik.dammert,hans.hellsten@saabgroup.com}
}
\and
H. ~Hellsten${}^{\ast\!\ast}$%
}

\title{%
Wavelength-Resolution SAR
Ground Scene Prediction Based on Image Stack}

\newcommand{\myabstract}{%
This paper presents
five
different
statistical methods
for ground scene prediction~(GSP)
in
wavelength-resolution
synthetic aperture radar~(SAR) images.
The GSP image can be used as a reference image in
a change detection algorithm
yielding a high probability of detection
and low false alarm rate.
The predictions
are based on
image stacks,
which are
composed
of images
from the same scene acquired
at different instants
with the same flight geometry.
The considered methods for obtaining
the ground scene prediction
include
(i)~autoregressive models;
(ii)~trimmed mean;
(iii)~median;
(iv)~intensity mean;
and
(v)~mean.
It is expected that
the predicted image presents
the true ground scene without
change
and
preserves the ground backscattering pattern.
The study indicate that the
the median method provided the most
accurate representation of the true ground.
To show the applicability of the GSP,
a change detection algorithm was considered
using the median ground scene as a reference image.
As a result,
the median method
displayed the probability
of detection of~$97\%$
and a false alarm rate of~$0.11/\text{km}^2$,
when considering military vehicles concealed in a forest.
}

\newcommand{\mykeywords}{%
CARABAS~II;
ground scene prediction;
image stack;
multi-pass;
SAR images.
}

\date{}

\begin{document}

\printtitle

\section{Introduction}

Common
tasks
in synthetic aperture radar~(SAR)
statistical
image processing
include
the
identification
and
classification
of
distinct
ground type~\cite{hoekman2000l, tison2004,Cintra2013,inglada2007,Palm2019},
modeling~\cite{sportouche2017,eltoft1998,
amirmazlaghani2009,belcher2013},
and
change detection~\cite{gudnason2009,zheng2014,Ulander2005,mercier2008}.
In special,
wavelength-resolution
low-frequency SAR systems
are
useful for
natural disasters monitoring,
foliage-penetrating applications,
and detection
of concealed targets~\cite{ulander2004}.

The
wavelength-resolution SAR system is usually associated with
ultrawideband~(UWB)
radar signal
and ultrawidebeam antenna~\cite{hellsten1996}.
With such, the maximum resolution is achieved
and it is in the order of radar signal wavelength.
Additionally,
available
UWB SAR systems only
operate
at low frequencies.
One essential feature of
wavelength-resolution SAR systems
is that the speckle noise does not influence
the
acquired
images
since it is likely that only a single scatter
is present in the resolution cell.
Additionally,
small
scatterers present
in the ground area of
interest do not contribute
to the backscattering
for low-frequency radar systems.
Thus, small structures, such as tree
branches and leaves,
are not shown
in SAR images~\cite{Renato2016}.
Because
large scatterers
are associated
with low-frequency
components,
they tend to be
less influenced by
environmental effects and
are
stable in time.
Hence,
by using
multi-passes
with identical
heading and incidence angle
of the illuminating platform at a given ground area,
an image package with similar statistics
can be obtained~\cite{vu2017}.
In~\cite{dimas2019},
it is
discussed
clutter statistical models
for stacks of  very-high-frequency~(VHF)
wavelength-resolution SAR images.
The SAR image stacks
are a frequent topic
of study
for SAR systems with high resolution~\cite{baselice2013markovian,
montazeri2016three,wang2014efficient}.
However,
the literature lacks
the use of large image stacks for wavelength-resolution SAR
for change detection applications.

Change detection algorithms~(CDA)
have been widely considered over the years
in the detection
of distinct targets
in SAR images~\cite{white1991,ulander2001,vu2018}.
In particular,
the
wavelength-resolution
SAR
change detection
is an
important topic of research
and has been
studied
for more than a decade~\cite{vu2017}.
Wavelength-resolution
systems have also shown
unique results
with high detectability
rate on low false alarm rate per square km,
as presented, for exmaple,
in~\cite{vu2017,vu2018}.
The nature of the
wavelength-resolution SAR imagery
can be exploited
to facilitate the design of CDAs,
since
(i)~the contribution
of small scatterers to radar echoes is not significant
for the wavelength of several meters;
(ii)~scatter
from large objects
are the main contribution;
(iii)~large scatterers are usually stable in time
and less sensitive to environmental effects;
and
(iv)~the wavelength-resolution
almost totally cancel the speckle noise~\cite{Renato2016}
in the SAR image given
a very stable backscattering between measurements.

A CDA is
used to
detect changes in a ground scene between distinct
measurements
in time,
such as
natural disasters like floods and wildfires
or human-made interferences~\cite{ulander2004}.
Generally, in wavelength-resolution systems,
a CDA
can be simply obtained by the
subtraction of two single-look images
(reference and surveillance),
followed by a thresholding operation.
However,
an image stack
can be considered instead of just two images
in a CDA;
such a collection of images leads
to improved detection performance,
as discussed in~\cite{vu2017}.
This information
is used to eliminate clutter and noise
in the surveillance image~\cite{vu2017},
and consequently,
enhancing CDA results.
Recently, a study using a small stack of
multi-pass
wavelength-resolution
SAR images for
change detection was introduced in~\cite{vu2017}.

In~\cite{Palm2018},
the autoregressive~(AR) model
was employed
as a preliminary study
considering a
ground scene prediction~(GSP)
based on
a single
wavelength-resolution SAR
image stack.
The resulting predicted image was submitted
as input data
to a change detection algorithm,
based only on
subtraction,
thresholding,
and morphological operations.
The~CDA in~\cite{Palm2018}
corresponds
to the
detection analysis step
of the~CDA
used in~\cite{Lundberg2006}.
Despite its simplicity,
the change detection results
in~\cite{Palm2018}
were
competitive when compared with
the ones recently presented in~\cite{Viet2017,
vu2017}.

Multi-pass SAR images
cannot be exactly equidistantly observed
over time
since
the
noise across the image stack
is not related to the time order.
As a consequence,
the use of a time series model,
commonly
employed
in
statistical signal processing~\cite{ghirmai2015,liu2014,
biscainho2004,milenkovic1986},
may not be the most
suitable
approach
to obtain a GSP,
and consequently, resulting in lower performance in
a CDA.
Additionally,
the backscattering
of the images
in the stack
is stable in time,
i.e.,
a sequence of pixels for each position
follows a similar pattern,
and changes in such behavior
are understood
as outliers.
Thus,
an
image filtering
considering
robust statistical
methods,
such as trimmed mean
and median~\cite{robust,hampel2011},
might be better
candidates
to obtain a ground scene prediction.
These
approaches
can provide an accurate
prediction of the ground scene,
avoid the time order problem,
and exclude the pixels that do not
follow the sequence pattern.
Indeed,
the median and the trimmed mean
filters are traditionally used to
remove impulse noise from an image~\cite{bustos2009,
wang1999progressive,
kirchner2010detection,chen2013blind,zhang2014revealing,
chen2015median,
oten2004adaptive,ahmed2013removal}.

To the best
of our knowledge,
the study in~\cite{Palm2018}
is the only work
related to the ground scene prediction for
wavelength-resolution SAR image stacks.
Our paper
extends the
results
presented in~\cite{Palm2018}
with four other statistical
methods to predict a
ground scene for three SAR image stacks,
since
statistical methods
are commonly employed
in
SAR image processing~\cite{hoekman2000l, tison2004,
gudnason2009,zheng2014,mercier2008,eltoft1998,
amirmazlaghani2009,
belcher2013,
sportouche2017,Palm2019}.
The selected
statistical
methods
to obtain the
prediction image
are
(i)~autoregressive models;
(ii)~trimmed mean;
(iii)~median;
(iv)~intensity mean;
and
(v)~mean.
The
predicted ground scene
methods
are sought
to
preserve the
ground backscattering statistical
characteristics of the images
in the stack and presents
predicted pixel values closer to the
original images.
It is expected that
the predicted images
represent
the true ground scenes,
allowing applications, such as monitoring
of forested areas and natural disasters.
In this paper, our goal is twofold.
First, we propose
the use of statistical methods
to obtain a ground scene prediction
image
based on
a
wavelength-resolution SAR image stacks.
Second,
we consider this new image
as a reference image
in a change detection algorithm.
In particular,
we employed
the
median GSP image
obtained
based on stack statistics as a reference image
in a
CDA
based on the
detection analysis step
of the
CDA presented in~\cite{Lundberg2006},
which was evaluated in terms
of target detection probability
and false alarm rate.
The results
reported in~\cite{vu2017,Ulander2005,vu2018}
were adopted
as the reference
model
for comparison.

The paper is organized as follows.
In Section~\ref{s:cda},
we describe
the considered
change detection method
and
a suite of selected statistical methods
for ground scene prediction.
Section~\ref{s:resul}
presents
experimental results,
including
a description
of the considered data set,
the
ground scene prediction
results,
and
the change detection results.
Then, a change detection method
based on the discussed GSP approaches
is introduced.
Finally,
Section~\ref{s:concl}
concludes the paper.

\section{Change Detection Method}
\label{s:cda}

\begin{figure}
\centering
\includegraphics[scale=0.3]{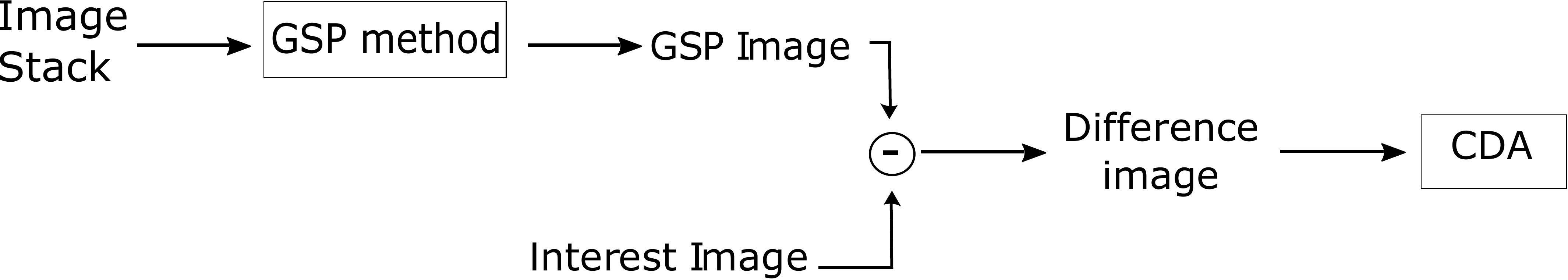}
\caption{Processing scheme for change detection.
The GSP image is the reference image
and the interest image
is the surveillance image.
The~CDA is performed
applying
thresholding and
morphological operations
in the difference image.
Note that the difference image
is based on the subtraction
between single-look image pixels
as a consequence
of the stability in backscattering using wavelength-resolution SAR system.
}
\label{f:proc}
\end{figure}

The change detection
method used in this paper
applied
the processing scheme given in Figure~\ref{f:proc}.
An image stack is processed by a desirable~GSP
method furnishing the~GSP image.
The changes
are simply obtained with
the
subtraction
of the image of interest~(surveillance image)
from the GSP image~(reference image).
For change detection,
we applied
thresholding to the difference
image and then used
morphological operations for false alarm minimization.
The methods employed to obtain
the GSP images
are described in the next section.

The employed CDA
consists of two
mathematical morphology
steps.
First,
an opening
operation~\cite{gonzalez2008}
aimed at removing small
pixel values,
which are regarded as noise.
The second step is a dilation
that prevents
the
splitting
of the interest targets
in multiple
substructures.
The first step uses a
$3 \times  3~\text{pixel}$
square structuring element,
whose size is determined by the
system resolution;
the second step
considers a
$7 \times  7~\text{pixel}$
structuring element,
which is linked
to the approximate size
of the targets~(about $10 \times  10~\text{pixels}$).

\subsection{Ground Scene Prediction}
\label{s:ge}

\begin{figure}
\centering
\includegraphics[width=0.7\textwidth]{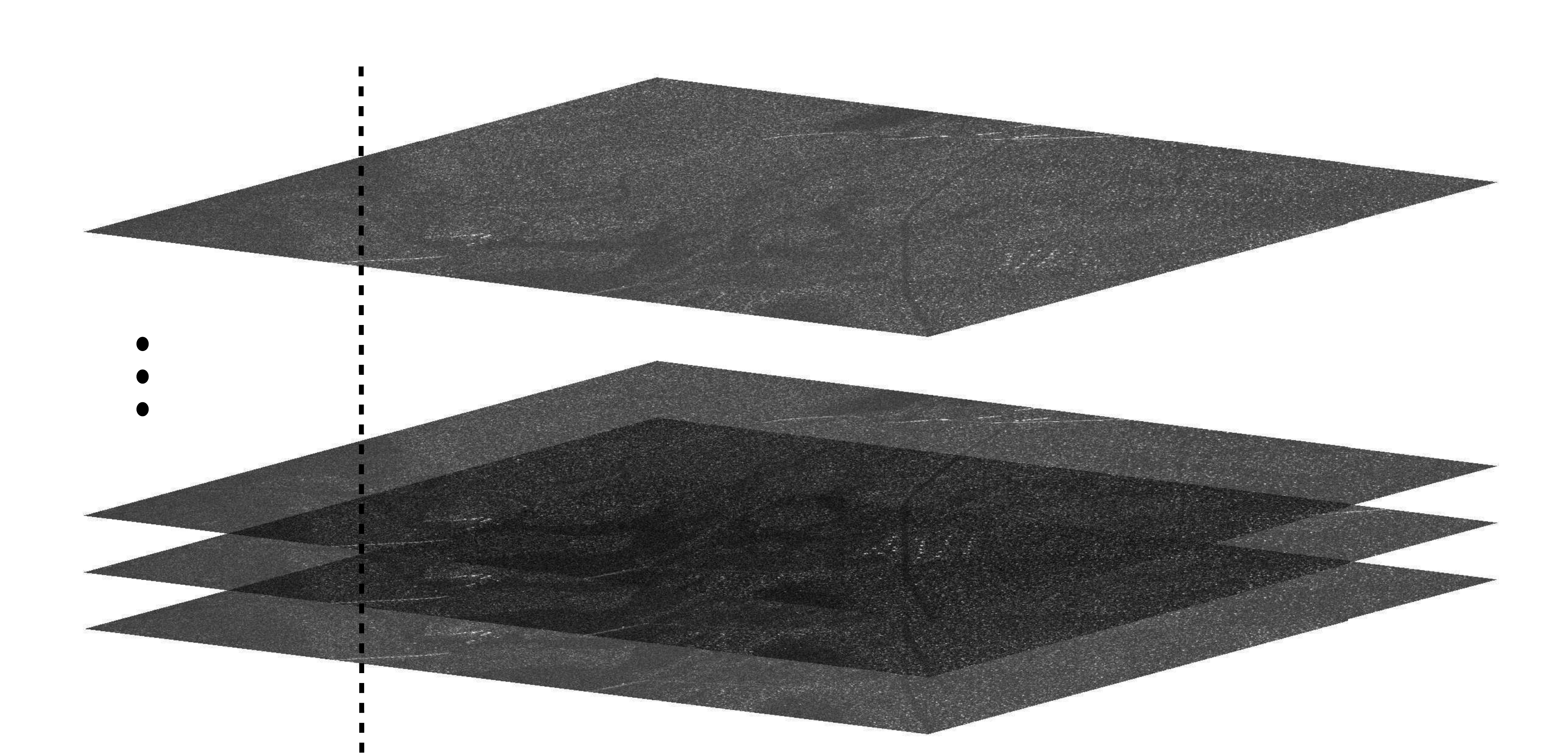}
\caption{Stack of images to be considered in~GSP.
The methods should be applied
for each pixel position, as evidenced
by the vertical line.}
\label{f:stack}
\end{figure}

As discussed in~\cite{dimas2019},
an image stack
is
composed of
images
with similar heading
and incidence angle
of the same
illuminating platform.
As a consequence of this similarity,
the SAR images in the stack
are very similar
and stable in time.
Thus,
a sequence
of each pixel position
can be extracted from the stack,
as illustrated in
Figure~\ref{f:stack}.

The data set considered in this paper is composed of
wavelength-resolution SAR images,
i.e.,
the resolution of the SAR image
is in the order of the radar signal wavelength~\cite{Renato2016}.
Therefore, there may only be a single scatter
in the resolution cell.
As a consequence,
the considered images
are not
affected by speckle noise,
which is
typically a strong source of noise in
SAR images in higher frequency bands.
Thus,
the backscattering
from the image stack
is
stable in time,
allowing an accurate~GSP.

We consider
five statistical
methods to obtain
ground scene predictions.
The techniques
are applied in a sequence
of pixels,
as
described
in the following.

\subsection{AR Model}

The
AR model was adopted
to compute the~GSP,
which
can be defined as~\cite{Kay1998-2}
\begin{align}
\label{e:model}
y[n] = - \sum_{k=1}^p a[k] y[n-k] + u[n]
,
\quad
n = 1, 2, \ldots, N
,
\end{align}
where~$y[n]$ is the
value
of each pixel in one image,
$N$ is the number of images in the stack,
$a[k]$~are the autoregressive terms,
$u[n]$~is white noise,
and~$p$ is the order of the model~\cite{Kay1998-2}.
The autoregressive terms~$a[k]$
in~\eqref{e:model}
can be estimated by the Yule-Walker method~\cite{Brockwell2016,Kay1998-2}.

Hence, the estimated autoregressive
terms~$\widehat{a}[k]$
are the solutions of the equation system,
given by~\cite{Kay1998-2}
\begin{align}
\label{e:yw}
\begin{split}
\begin{bmatrix}
r_{yy}[0] & r_{yy}[1] & \ldots & r_{yy}[p-1] \\
r_{yy}[1] & r_{yy}[0] & \ldots & r_{yy}[p-2] \\
\vdots & \vdots & \ddots & \vdots \\
r_{yy}[p-1] & r_{yy}[p-2] & \ldots & r_{yy}[0]
\end{bmatrix}
\begin{bmatrix}
a[1] \\ a[2] \\ \vdots \\ a[p]
\end{bmatrix}
= -
\begin{bmatrix}
r_{yy}[1] \\ r_{yy}[2] \\ \vdots \\ r_{yy}[p]
\end{bmatrix}
,
\end{split}
\end{align}
where~$r_{yy}[\cdot]$ is the sample autocorrelation
function.
Information about large
sample distributions of the Yule-Walker
estimator, order selection, and confidence
regions for the
coefficients can be found in~\cite{Brockwell2013}.
Considering the
estimated autoregressive terms~$\widehat{a}[k]$,
it is possible to forecast~$h$ steps
ahead
with the AR model as~\cite{Brockwell2016}
\begin{align}
\widehat{y}[N+h] = - \sum \limits _{k=1} ^p
\widehat{a}[k] y[N+h-k]
.
\end{align}
The ground scene prediction image
is obtained by forecasting
the one-step ahead~($h=1$)
pixel value for each pixel
in the image.

\subsection{Trimmed Mean, Median, and Mean}

For SAR images
whose backscattering is stable in time,
robust methods
can be applied
to obtain a~GSP.
We consider the
trimmed mean
to obtain a~GSP,
which
is given by
\begin{align}
\bar{y}_{\text{tm}}
=
\frac{2}{N - 2m}
\sum \limits _{n=m+1}^{N-m}
y^\star[n]
,
\end{align}
where $y^\star[n]$
is the ordered sequence of
$y[n]$,
$m=(N-1)\alpha$,
and~$\alpha \in [0,1/2)$~\cite{robust,hampel2011}.
If~$\alpha = 0$
or~$\alpha \rightarrow 0.5$,
then
the trimmed mean
corresponds
to the sample mean
and median,
respectively~\cite{robust},
which are
considered as methods for~GSP
derivation.

\subsection{Intensity Mean}

We also use
the intensity mean
for obtaining
ground scene
predictions,
given by
\begin{align}
\bar{y}_{\text{im}}
=
\sqrt{
\frac{1}{N}
\sum \limits _{n=1}^{N}
y[n] ^2
}
.
\end{align}
Compared to other
statistical methods,
the
intensity mean has
the advantage
of providing
physical
interpretation
about the image reflection.
However,
the intensities
values
contribute
evenly
to
the prediction results,
which can be
strongly
affected
by the changes
in the ground scene~\cite{robust}.

\section{Experimental Results}
\label{s:resul}

In this section,
we present the results
obtained
from
the discussed ground scene
prediction methods
and
describe
an approach
for
change detection
based on such methods.

\subsection{Data Description}

In this study,
we considered a data set
obtained from
CARABAS~II, a Swedish
UWB VHF SAR
system whose images are available
in~\cite{data}.
The system is a low-frequency
wavelength-resolution system
which means that the images have
almost no speckle noise.
The data set
was
divided into three stacks
with eight images
each,
i.e.,
two out of six passes have identical
flight headings.
Two passes have a flight heading
of~$255^{\circ}$,
two of~$135^{\circ}$,
and two of~$230^{\circ}$,
and
the heading is defined
as~$0^{\circ}$ pointing towards
the north with clock-wise increasing heading.
The images in the stacks
have the same flight geometry but are associated
with four different targets deployments (missions
1 to 4) in the ground scene.
Hence,
with four missions and six passes for each mission,
there is~24 magnitude single-look SAR images.
The images cover
a scene
of size
$2 \, \text{km} \times 3 \, \text{km}$
and
are georeferenced
to the Swedish reference system RR92,
that can easily be transformed to WGS84~\cite{Ulander2005,Lundberg2006}.

The first stack is composed of images
corresponding to flight passes~1 and~3;
the second stack,
with passes~2 and~4;
and the last stack is composed of images
associated with passes~5 and~6.
In all images, the backscattering was stable in time,
and
only target changes
are expected
within the image stacks.

\begin{figure}
\centering
\includegraphics[scale=0.35]{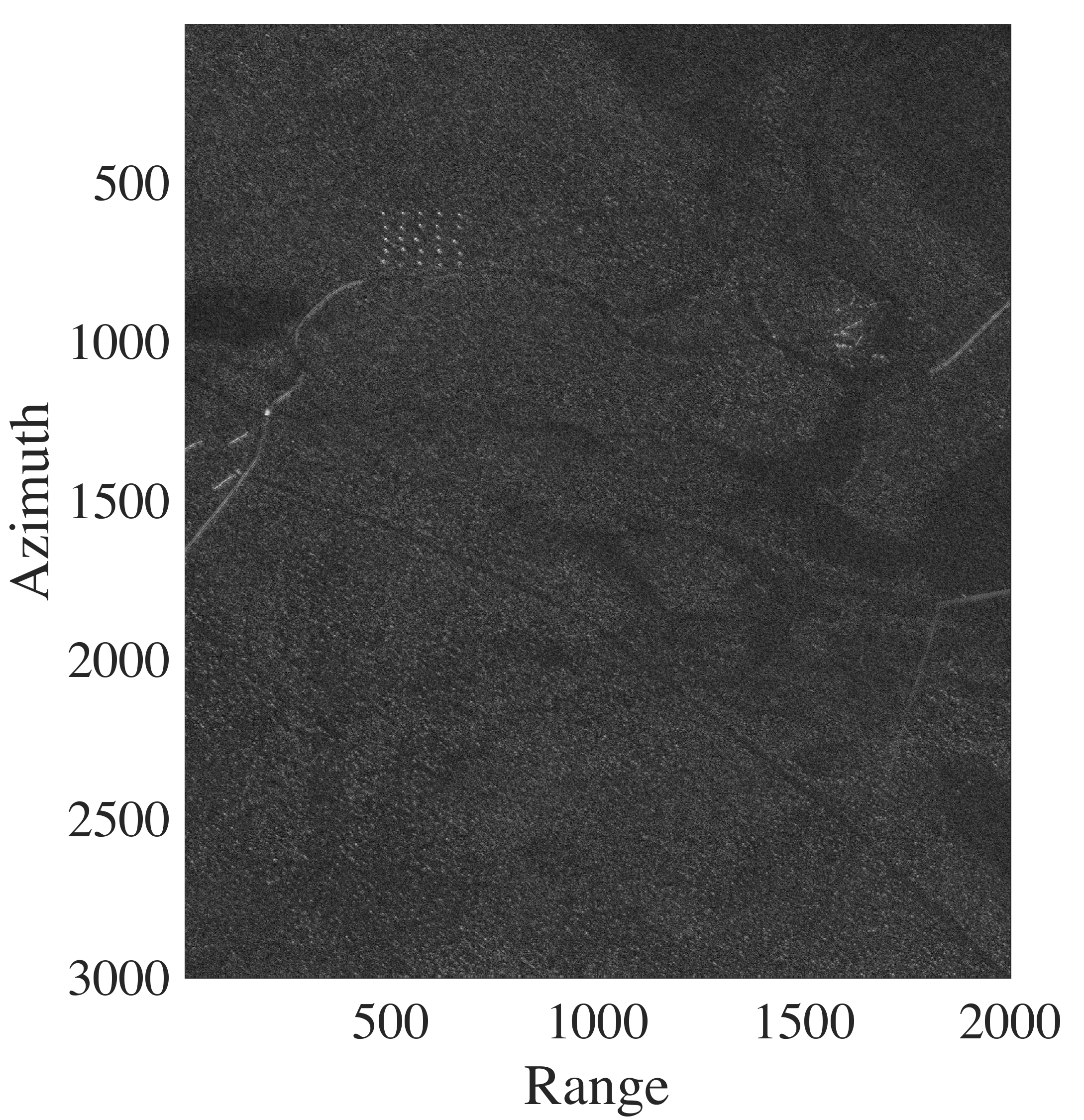}
\caption{Sample image from
	CARABAS~II data set---Stack 1 -- mission 1 and pass 1.
	}
\label{f:fig}
\end{figure}

Each image is
represented as
a matrix of
$3000 \times 2000$
pixels, corresponding
to
an area of $6~\text{km}^2$.
As reported in~\cite{Ulander2005},
the  spatial resolution of CARABAS II  is
$2.5~\text{m}$ in azimuth
and
$2.5~\text{m}$ in range.
The ground scene is
dominated by boreal forest with pine trees.
Fences, power lines, and roads were also present in the scene.
Military vehicles
were deployed in the SAR scene and
placed uniformly,
in a manner to facilitate
their identifications in the tests~\cite{Lundberg2006}.
Each image
has~$25$ targets
with
three
different
sizes
and
the spacing
between the vehicles was about 50 meters.
For illustration,
one image
of Stack~1
is shown in Figure~\ref{f:fig}.
In this image,
the vehicles were
(i)~obscured by foliage;
(ii)~deployed  in the top left of the scene;
and
(iii)~oriented in a south-western heading.
This deployment corresponds
to mission 1.
In missions 2, 3, and 4,
these vehicles
were deployed
in other locations
and
were oriented in a north-western,
south-western,
and
western heading, respectively~\cite{Ulander2005,Lundberg2006}.

\subsection{Ground Scene Prediction Evaluation}

\begin{figure*}
	\centering
	\subfigure[AR model]
{\includegraphics[scale=0.25]{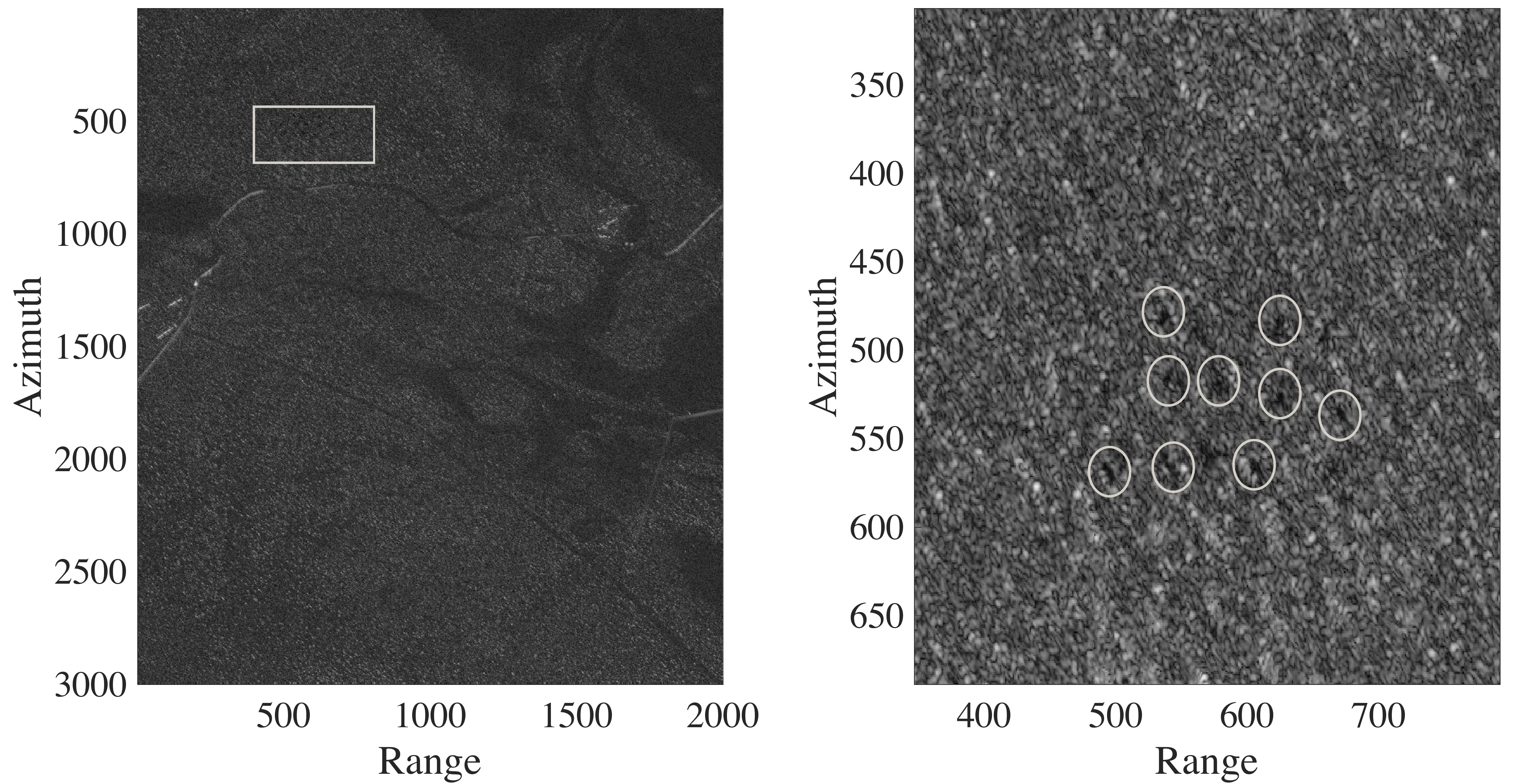}}
	\centering
	\subfigure[Mean]
	{\includegraphics[scale=0.25]{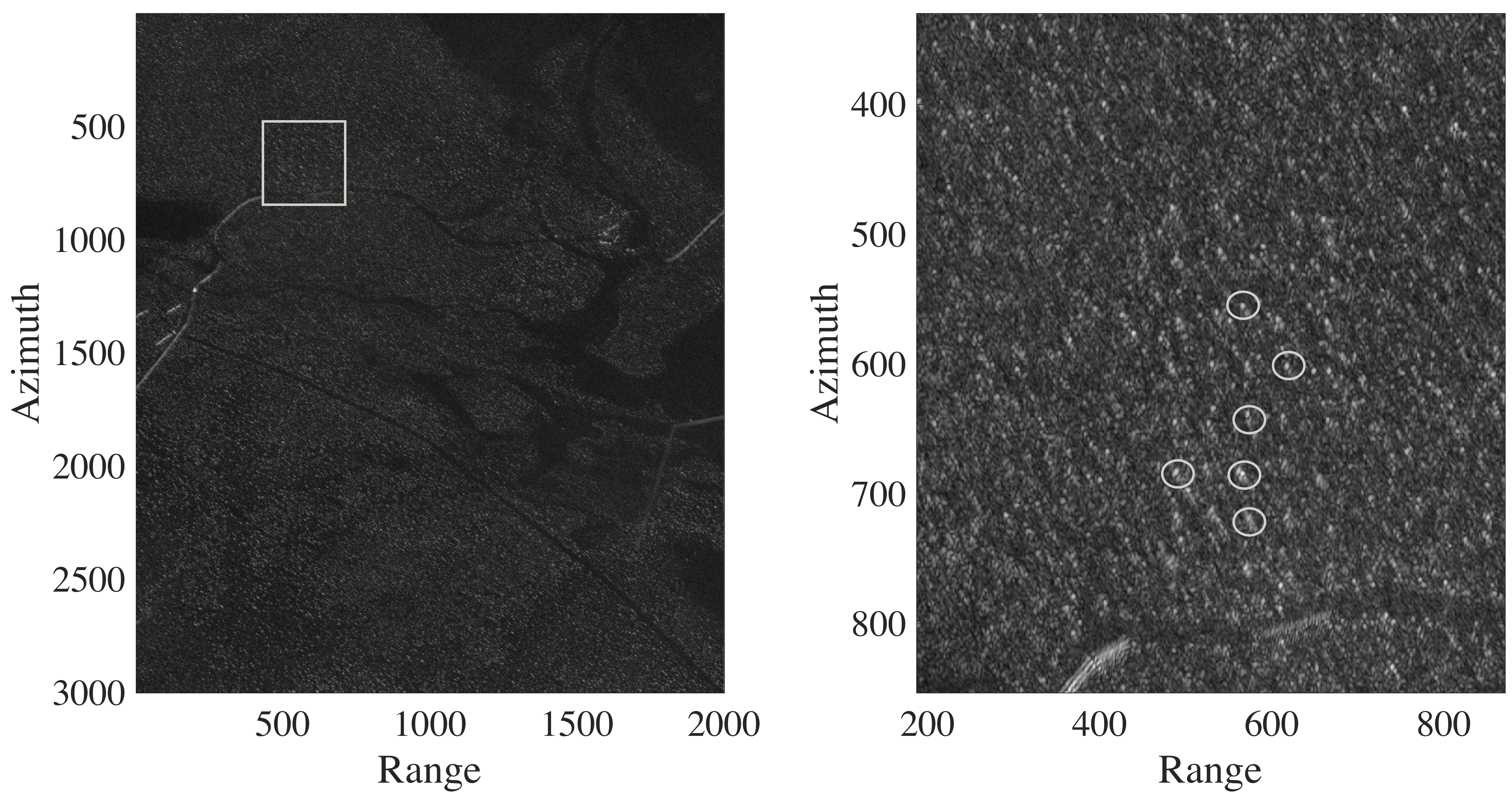}}
	\centering
\subfigure[Intensity mean]
{\includegraphics[scale=0.25]{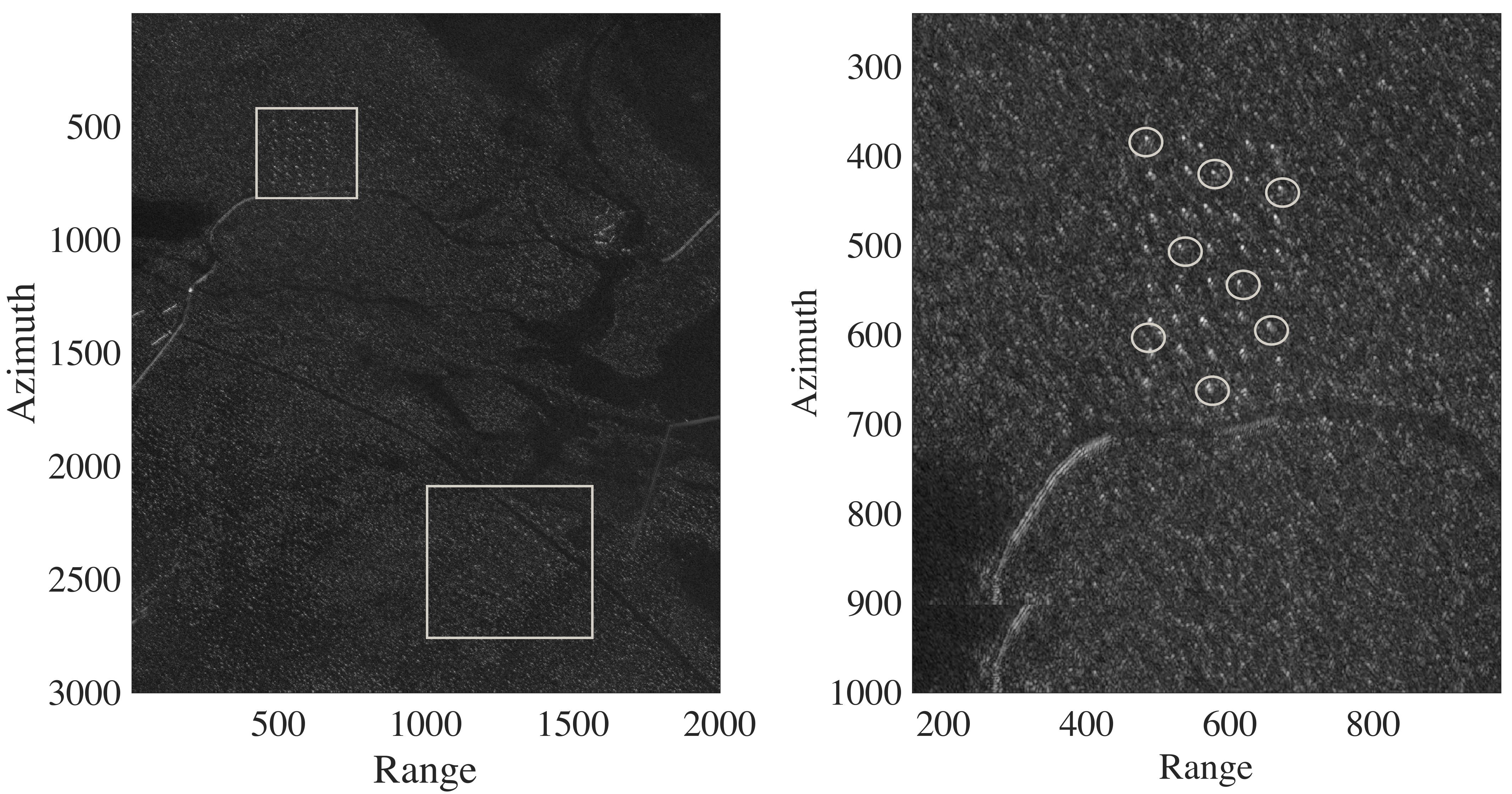}}

	\caption{Ground scene prediction images
		for Stack~1
		based on
		AR model,
		mean,  and intensity mean
		methods.
		The areas highlighted by
		rectangles in the images
		represent the regions
		where the targets are deployed.
		The circles show
		selected military vehicles that can be viewed.
		}
	\label{f:gsp1}
\end{figure*}

\begin{figure*}
	\centering

\subfigure[Trimmed mean]
{\includegraphics[scale=0.3]{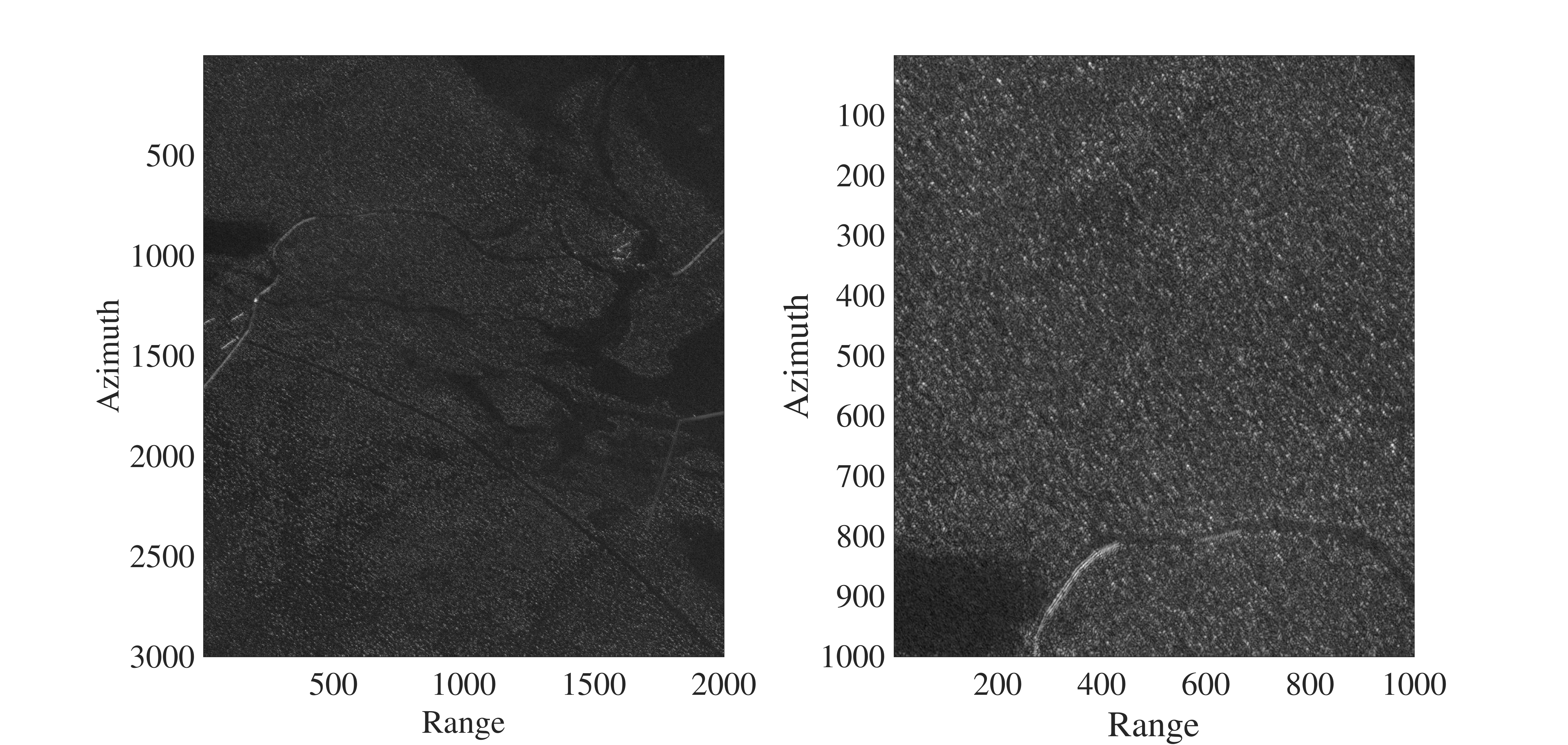}}

\subfigure[Median]
{\includegraphics[scale=0.3]{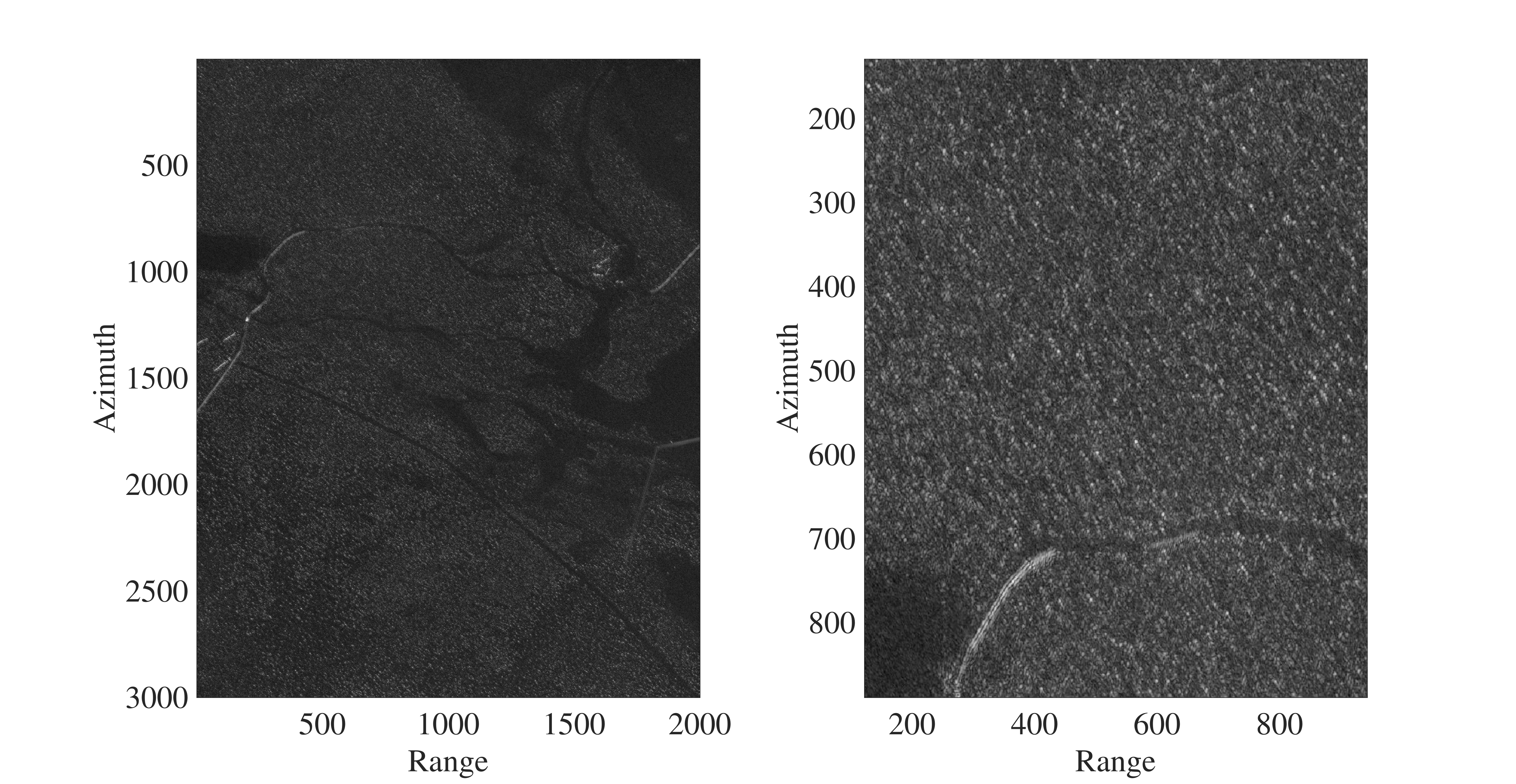}}
	\caption{Ground scene prediction images
		for Stack~1
		based on
		trimmed mean and median methods.
		}
	\label{f:gsp2}
\end{figure*}

The
AR model
parameter
estimation
requires
(i)~fitting 6,000,000 models
(one fit for each pixel)
in each stack
and
(ii)~evaluating the best model for each pixel sequence.
Such demands lead to a significant computational burden.
For
simplicity,
we considered~$p=1$
in the AR model.
Within the image stack,
the
two images
related to the targets
have the highest pixel values
in the areas where the targets were deployed.
Thus,
to compute the trimmed mean,
we
considered
$m = 2~(\alpha \approx 0.3)$,
expecting
to
remove the pixels
related to the targets,
since
it is
desired that the predicted image
presents the true ground scene without change.
Figures~\ref{f:gsp1}
and~~\ref{f:gsp2}
show the ground scene prediction for Stack~1,
considering the discussed methods
and a zoomed image in the region
where the targets
were deployed.
In Figure~\ref{f:gsp1},
the deployed targets
are visually present.
However,
the targets
are
absent in
the images predicted with the
trimmed mean
and
median,
as shown in
Figure~\ref{f:gsp2}.
The areas highlighted by
rectangles and circles in the images
in Figure~\ref{f:gsp1}
indicate the regions where the targets
were deployed during the measurement campaign.
The circles show
selected military vehicles that can be viewed.
With
such visual analysis, the
trimmed mean
and
median show better performance, i.e.,
better prediction of the ground scene.
For brevity,
we limited our presentation
to the~GSP images
from Stack~1,
which is representative
of all considered stacks.

Table~\ref{t:results}
displays descriptive statistics
of the employed images,
such as
average, standard deviation,
skewness, and kurtosis.
It is desirable that
a~GSP presents not only
a good visual
representation of the true
ground but also preserves the
statistical characteristics of the
image of interest.
In Table~\ref{t:results},
we
highlighted
the two best
methods
according to each
considered measure.
In
the majority
of the scenarios,
the AR model and
median methods
outperformed
the
remaining methods.

\begin{table}
\centering
\caption{Average, standard deviation, skewness,
and kurtosis
of one interest image and the ground scene prediction.
The interest image in Stack 1, 2, and 3,
is the image
of mission 1 and pass 1, 2, and 5,
respectively.
The two values of each measure that
yielded the closest
values with the
interest image
are
highlighted}
\label{t:results}
\begin{tabular}{ccccccc}
\toprule
 & Average & Standard  & Skewness & Kurtosis \\
 & & deviation & & \\
\midrule
\multicolumn{5}{c}{Stack 1} \\
\midrule
Interest image & $ 0.1442$ & $0.0894$ & $1.8597$ & $14.1740$ \\
AR model & $0.1101$ & $\textbf{0.0725}$ & $\textbf{2.1120}$ & $\textbf{13.5190}$ \\
Trimmed mean & $\textbf{0.1430}$ & $0.0680$ & $2.9051$ & $21.2919$ \\
Median & $\textbf{0.1424}$ & $\textbf{0.0688}$ & $\textbf{2.8231}$ & $\textbf{20.4990}$  \\
Mean & $0.1467$ & $0.0663$ & $3.0516$ & $22.8448$ \\
Intensity mean & $0.1592$ & $0.0667$ & $3.0090$ & $22.8725$ \\
\midrule
\multicolumn{5}{c}{Stack 2} \\
\midrule
Interest image & $0.1373$ & $0.0968$ & $2.9345$ & $30.5666$ \\
AR model & $0.0997$ & $0.0784$ & $\textbf{3.6398}$ &  $\textbf{40.9991}$ \\
Trimmed mean & $\textbf{0.1344}$ & $\textbf{0.0806}$ & $4.4488$ & $55.4260$  \\
Median & $0.1339$ & $\textbf{0.0812}$ & $\textbf{4.3664}$ & $\textbf{53.9367}$   \\
Mean & $\textbf{0.1376}$ & $0.0792$ & $4.6022$ & $58.3558$  \\
Intensity mean & $0.1485$ & $0.0792$ & $4.5487$ & $ 57.8894$  \\
\midrule
\multicolumn{5}{c}{Stack 3} \\
\midrule
Interest image & $0.1451$ & $ 0.0905$ & $1.8583$ & $14.0932$ \\
AR model & $0.0997$ & $\textbf{0.0683}$ & $\textbf{2.2034}$ & $\textbf{14.6539}$  \\
Trimmed mean & $ \textbf{0.1372}$ & $0.0665$ & $2.8811$ & $22.0954$ \\
Median & $0.1366$ & $\textbf{0.0674}$ & $\textbf{2.8090}$ & $\textbf{21.3242}$  \\
Mean & $\textbf{0.1410}$ & $0.0646$ & $2.9582$ & $22.9540$ \\
Intensity mean & $0.1534$ & $0.0655$ & $2.9170$ & $ 22.9794$  \\
\bottomrule
\end{tabular}
\end{table}

To evaluate the difference
between the ground scene prediction
methods,
we computed some standard quality
adjustment measures.
The criteria are the mean square
error~(MSE),
mean absolute percentage error~(MAPE),
and
median absolute error~(MdAE),
which can be defined as follows~\cite{hyndman2006}

\begin{align}
{\textrm {MSE}}
& =
\frac{1}{Q}\sum \limits _{q=1}^Q (x[q]-\widehat{x}[q])^2
,
\\
{\textrm {MAPE}}
&=
\frac{1}{Q}\sum \limits _{q=1}^Q
\frac{\vert x[q]-\widehat{x}[q]\vert}{\vert x[q] \vert}
,
\\
{\textrm {MdAE}}
& =
\operatorname{Median}
\left(
\vert x[q]-\widehat{x}[q] \vert
\right)
,
\quad
q=1,2,\ldots,Q
,
\end{align}
where
$x[q]$
and
$\widehat{x}[q]$
are the pixel values
of the interest
and
predicted
images
respectively,
$Q$ is the
number
of pixels,
and~$\operatorname{Median}(\cdot)$
is the median value
of~$\vert x[q]-\widehat{x}[q] \vert$,
for~$q=1,2,\ldots,Q$.
These
goodness-of-fit measures
are usually
considered to compare different
methods applied
to the same data set~\cite{hyndman2006}.
They
are expected to be
as close to zero as possible.
For the quality adjustment measures,
the target regions in the image were
excluded since we expect to obtain
an accurate ground scene prediction,
and
no target deployment should influence
the
measurements.
Table~\ref{t:men} summarizes
the results
of
the quality adjustment measures
for
the five considered statistical methods,
and
the
best
measurements
are
highlighted.
The  mean
method
presents the best performance
according
to
MSE measurements,
while the median method
excels
in terms of
MAPE and~MdAE
measures
in all the stacks.
However, the~MSE values
obtained with
the mean and
median methods
are similar.
The results provided
in Tables~\ref{t:results}
and~\ref{t:men}
consider the same reference image
of each stack.
Regardless of the selected image,
the median method
presented good performance
according to~MAPE,~MdAE,
and statistics measures.

\begin{table}
\centering
\caption{
Measures of quality of the ground scene prediction image.
The interest image in Stack 1, 2, and 3
is the image
of mission 1 and pass 1, 2, and 5,
respectively.
We highlighted
the values of each quality adjustment measure that
yielded the smallest values}
\label{t:men}
\begin{tabular}{cccccccc}
\toprule
& &  MSE & MAPE  & MdAE\\
\midrule
\multirow{6}{*}{Stack 1}
& AR model &  $ 0.0077$ & $0.6756$ & $0.0548$  \\
& Trimmed mean & $\textbf{0.0036}$ & $0.6187$ & $0.0364$ \\
& Median & $0.0037$ &  $\textbf{0.6125}$ & $\textbf{0.0351}$ \\
& Mean &  $0\textbf{.0036}$ & $ 0.6489$ & $0.0401$ \\
& Intensity mean & $0.0039$ & $0.7505$ &  $0.0426$ \\
\midrule
\multirow{6}{*}{Stack 2}
& AR model & $0.0068$ & $0.6450$ & $0.0502$ \\
& Trimmed mean & $\textbf{0.0030}$ & $0.5971$ & $0.0326$ \\
& Median &  $0.0031$ & $\textbf{0.5912}$ & $ \textbf{0.0315}$ \\
& Mean & $\textbf{0.0030}$ & $0.6254$ & $0.0359$ \\
& Intensity mean & $0.0032$ & $0.7204$ & $0.0378$ \\
\midrule
\multirow{6}{*}{Stack 3}
& AR model & $0.0083$ & $0.6337$ & $0.0557$ \\
& Trimmed mean & $0.0037$ & $0.5809$ & $0.0357$ \\
& Median & $0.0038$ & $\textbf{0.5751}$ & $\textbf{0.0346}$ \\
& Mean & $0\textbf{.0036}$ &  $0.6104$ & $0.0392$ \\
& Intensity mean & $0.0037$ & $0.7011$ & $0.0410$\\
\bottomrule
\end{tabular}
\end{table}

Based on visual inspection,
statistical characteristics,
and quality adjustment measures,
the median
method yields the most reliable prediction
among the considered
methods.
Therefore,
we separate the
predicted images
from
the
median method
as reference images
in the change detection algorithm
detailed in the next section.

\subsection{
Change Detection Results}

\begin{figure*}
	\centering

\subfigure[Pass one and mission one]
{\includegraphics[scale=0.35]{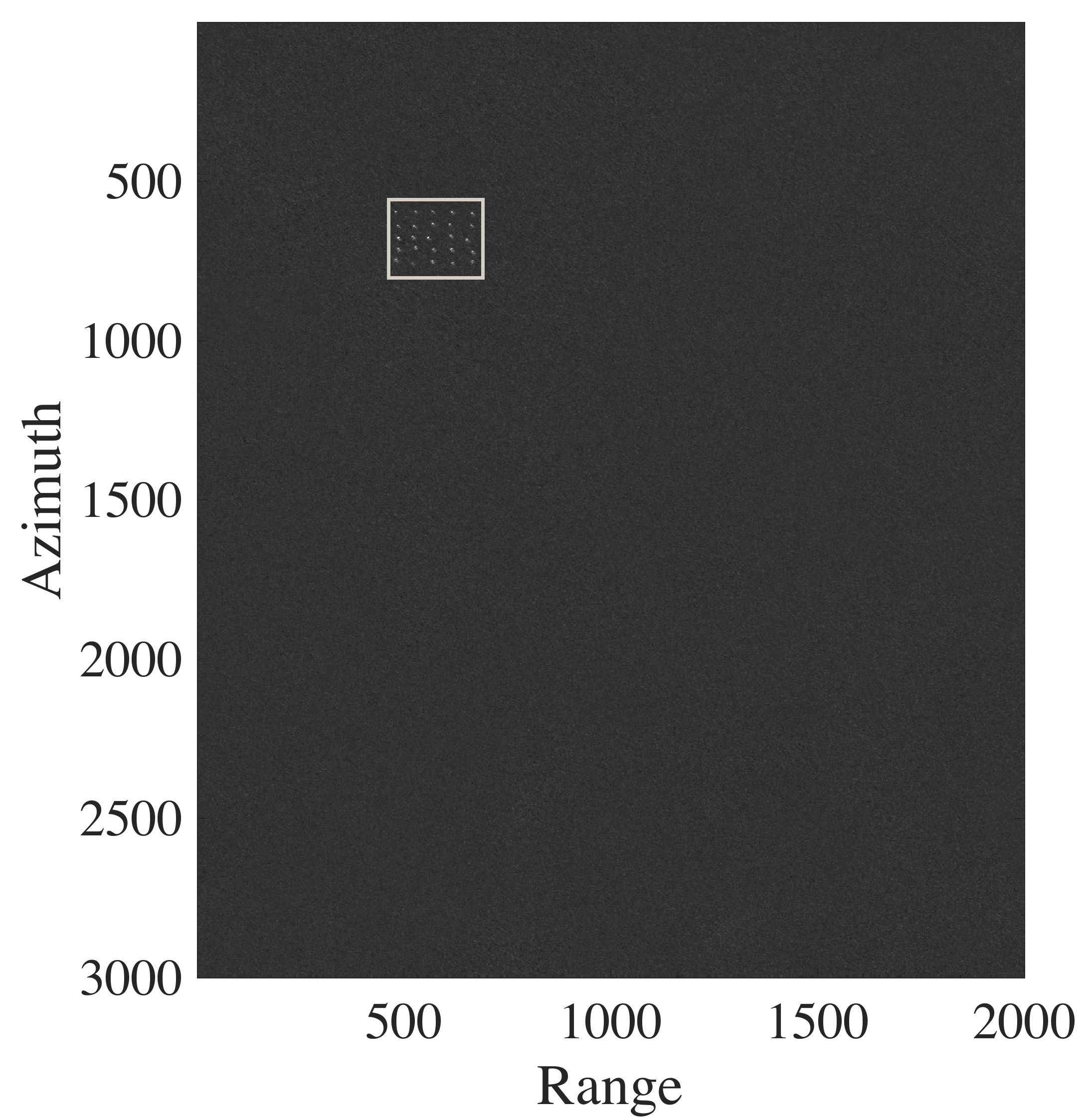}
\label{f:res1}}
\subfigure[Pass one and mission two]
{\includegraphics[scale=0.35]{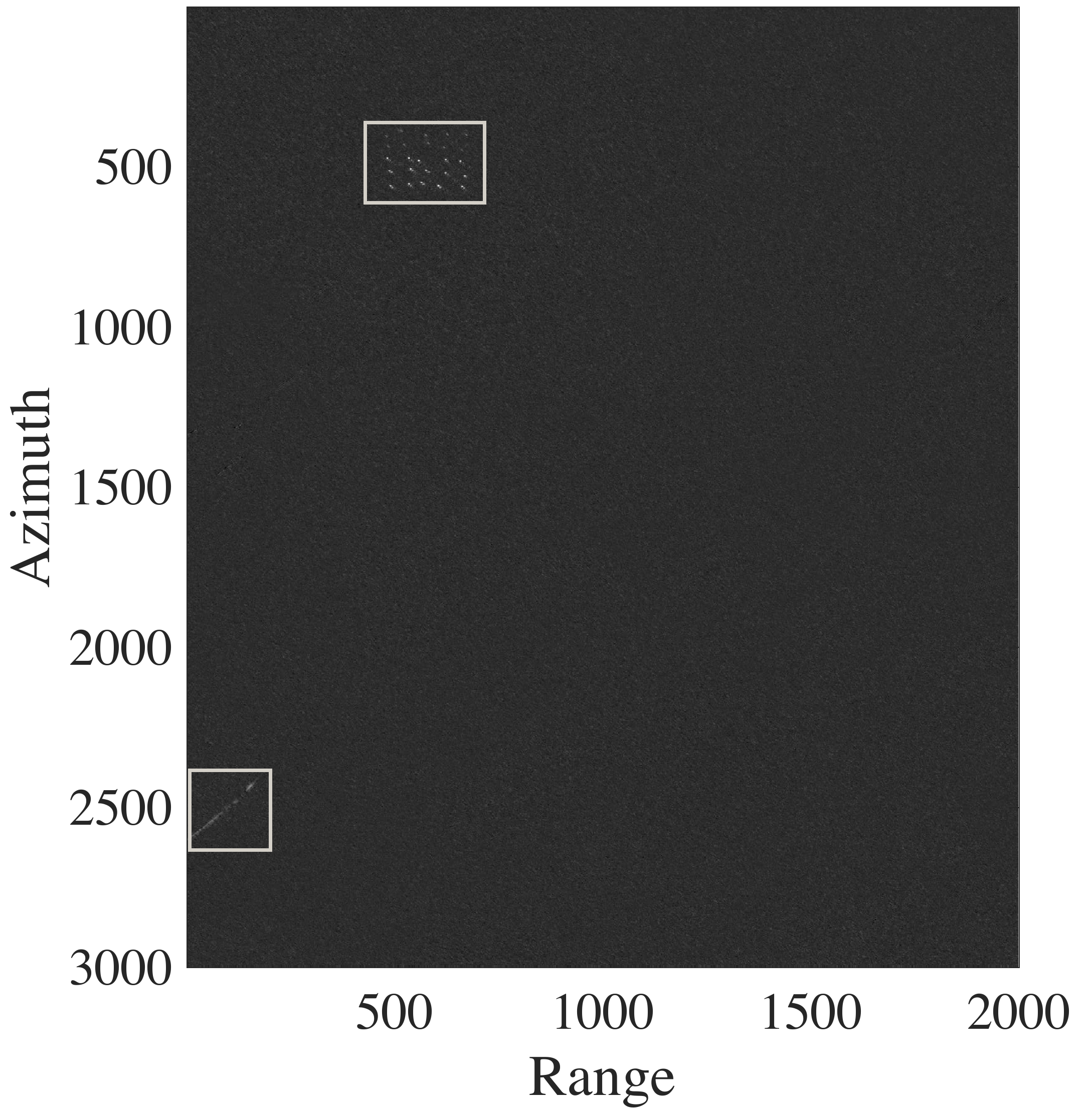}
\label{f:res2}}

	\caption{
	Subtraction of an interest image
	from the median ground scene prediction image.
	The areas highlighted
	by rectangles in the images represent
	the region with higher pixel values.
		}
	\label{f:res}
\end{figure*}

As indicated in Figure~\ref{f:proc},
we use the obtained
GSP image and the interest image
for change detection
based on image subtraction.
Two examples of
subtraction images
are shown in
Figure~\ref{f:res}.
Figure~\ref{f:res1}
highlights the deployed targets, while
Figure~\ref{f:res2}
focuses on the targets and the back-lobe structures.
A comparison
between
the
difference
image
shown in
Figure~\ref{f:res2}
to the
related~GSP image
suggests that
the
back-lobe
structures
are related
to
issues
in the SAR system
and image formation algorithm.

\begin{figure}
\centering
\includegraphics[scale=0.35]{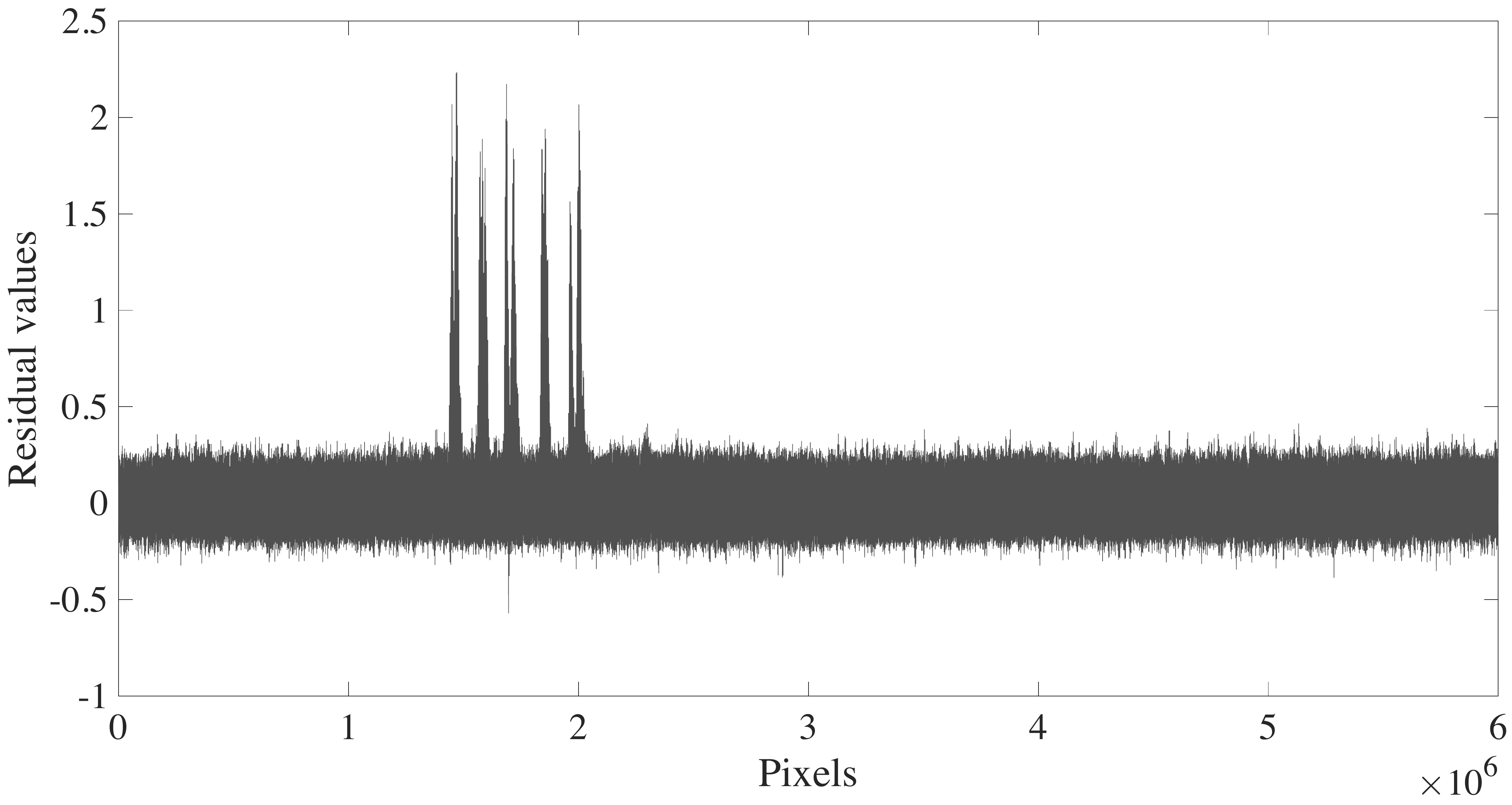}
\caption{Result of the
subtraction of
the ground scene prediction image
from the image obtained from mission 1 and pass 1.
}
\label{f:residual}
\end{figure}

Figure~\ref{f:residual}
shows the pixels values
of
the
image given in Figure~\ref{f:res1}
in a vectorized form.
In general,
the subtracted
image
pixels values
are randomly distributed in~$(-0.4,0.4)$.
As discussed in~\cite{Renato2016},
the distribution of the values
of the
CARABAS~II
subtracted image
approximately follows the Gaussian distribution
and the regions where no change occurs
are stable.
Thus,
the
threshold~($\lambda$)
can be simply chosen
as
\begin{align}
C = \frac{\lambda - \widehat{\mu}}{\widehat{\sigma}}
,
\end{align}
where~$C$ is a constant,
$\widehat{\mu}$~is the estimated mean,
and
$\widehat{\sigma}$
is
the
estimated
standard deviation of the
considered
amplitude
pixels
in the image.
For evaluation,
we set
$C \in \lbrace 2,3,4,5,6 \rbrace$,
resulting in
different false alarm rates (FAR),
which range from full detection
to almost null false alarm rate.

Table~\ref{t:detec}
summarizes the change detection results
corresponding to a single
constant~$C=5$.
Among 600 deployed vehicles
in the missions,~$579$
were
correctly detected.
There is~$22$
detected objects
that can not be
related to any vehicle
and are considered to be false alarms.
Thus,
the detection probability is about~$97\%$,
while the false alarm rate
is~$0.15/\text{km}^2$~(total of $144/\text{km}^2$).
Ten of
the~$22$
false alarms
are related to the back-lobe structures,
i.e., they are not actually false alarms
and
may stem from system and image formation issues.
Additionally,
in general,
the
undetected targets are related to
missions 2 and 4.
These
undetected  military vehicles
are more difficult to detect
since
they
have the smaller sizes
and
magnitude values,
and
consequently,
pixel values closer to
the forest ones.

\begin{table}
\centering
\caption{Change detection results obtained with~$C = 5$}
\label{t:detec}
\begin{tabular}{cccccccc}
\toprule
\multicolumn{2}{c}{Case of Interest} & Number of &
Detected & $\text{P}_d$ & 	Number of  \\
Mission & Pass & known targets & Targets &  & false alarms \\
\midrule
$	1	$	&	$	1	$	&	$	25	$	&	$	25	$	&	$	1.00	$	&		$	0	$		\\
$	2	$	&	$	1	$	&	$	25	$	&	$	25	$	&	$	1.00	$	&		$	3	$		\\
$	3	$	&	$	1	$	&	$	25	$	&	$	25	$	&	$	1.00	$	&		$	0	$		\\
$	4	$	&	$	1	$	&	$	25	$	&	$	23	$	&	$	0.92	$	&		$	2	$		\\
$	1	$	&	$	2	$	&	$	25	$	&	$	25	$	&	$	1.00	$	&		$	0	$		\\
$	2	$	&	$	2	$	&	$	25	$	&	$	25	$	&	$	1.00	$	&		$	1	$		\\
$	3	$	&	$	2	$	&	$	25	$	&	$	25	$	&	$	1.00	$	&		$	2	$		\\
$	4	$	&	$	2	$	&	$	25	$	&	$	23	$	&	$	0.92	$	&		$	1	$		\\
$	1	$	&	$	3	$	&	$	25	$	&	$	25	$	&	$	1.00	$	&		$	2	$		\\
$	2	$	&	$	3	$	&	$	25	$	&	$	23	$	&	$	0.92	$	&		$	0	$		\\
$	3	$	&	$	3	$	&	$	25	$	&	$	25	$	&	$	1.00	$	&		$	3	$		\\
$	4	$	&	$	3	$	&	$	25	$	&	$	23	$	&	$	0.92	$	&		$	0	$		\\
$	1	$	&	$	4	$	&	$	25	$	&	$	25	$	&	$	1.00	$	&		$	0	$		\\
$	2	$	&	$	4	$	&	$	25	$	&	$	25	$	&	$	1.00	$	&		$	0	$		\\
$	3	$	&	$	4	$	&	$	25	$	&	$	25	$	&	$	1.00	$	&		$	1	$		\\
$	4	$	&	$	4	$	&	$	25	$	&	$	23	$	&	$	0.92	$	&		$	0	$		\\
$	1	$	&	$	5	$	&	$	25	$	&	$	25	$	&	$	1.00	$	&		$	0	$		\\
$	2	$	&	$	5	$	&	$	25	$	&	$	15	$	&	$	0.60	$	&		$	6	$		\\
$	3	$	&	$	5	$	&	$	25	$	&	$	25	$	&	$	1.00	$	&		$	0	$		\\
$	4	$	&	$	5	$	&	$	25	$	&	$	24	$	&	$	0.96	$	&		$	0	$		\\
$	1	$	&	$	6	$	&	$	25	$	&	$	25	$	&	$	1.00	$	&		$	0	$		\\
$	2	$	&	$	6	$	&	$	25	$	&	$	25	$	&	$	1.00	$	&		$	1	$		\\
$	3	$	&	$	6	$	&	$	25	$	&	$	25	$	&	$	1.00	$	&		$	0	$		\\
$	4	$	&	$	6	$	&	$	25	$	&	$	25	$	&	$	1.00	$	&		$	0	$		\\
\midrule
\multicolumn{2}{c}{Total}	& $	600$ &	$579$	&	$0.97$	&		$22$	\\
\bottomrule
\end{tabular}
\end{table}

\subsection{
Evaluation}

\begin{figure}
\centering
\includegraphics[width=0.94 \textwidth]{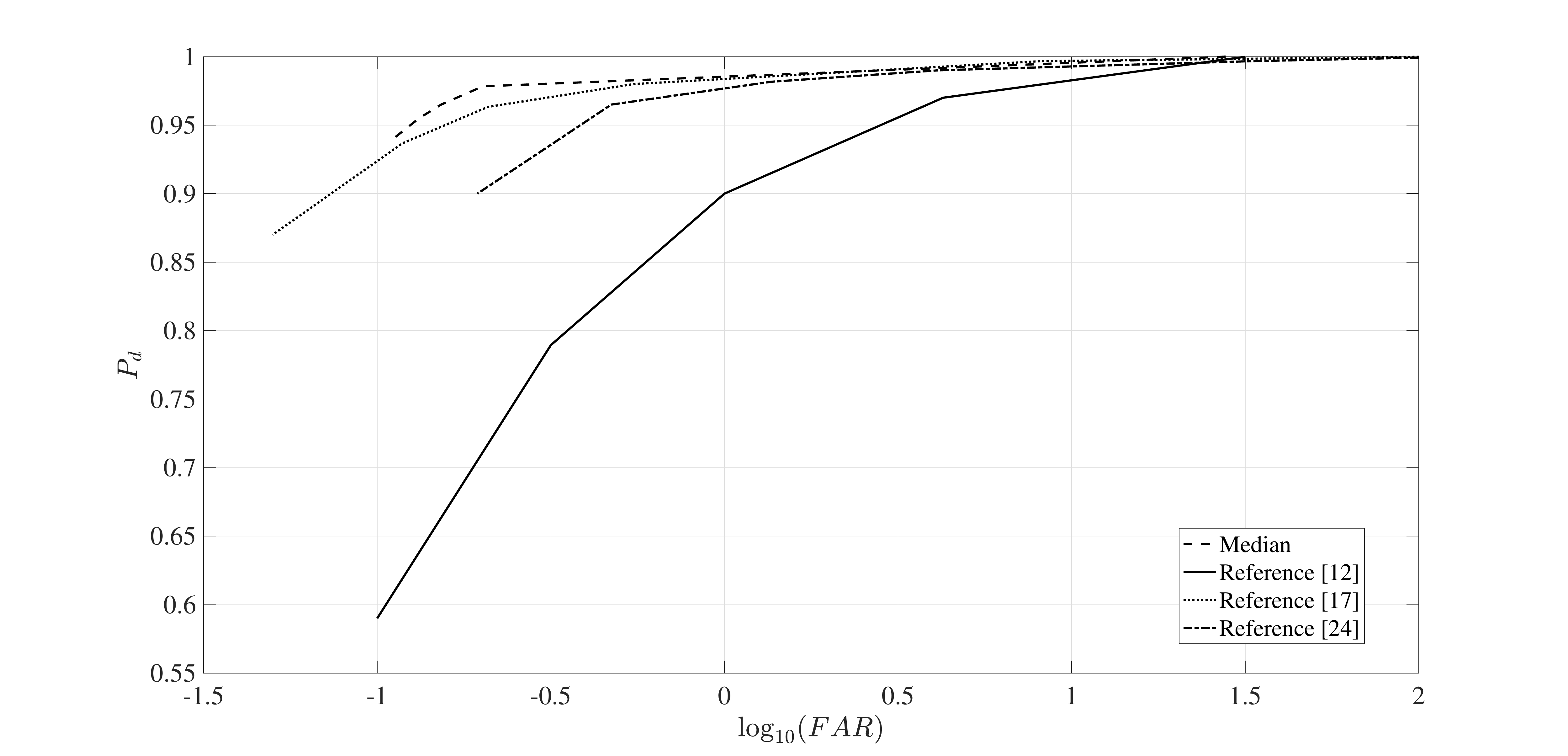}
\caption{
ROC curves obtained with the CDA with the background
predicted scene as the reference image
compared with the best ROC curves
extracted from~\cite{vu2017,Ulander2005,vu2018}.
}\label{f:roc}
\end{figure}

The performance of
change detection was evaluated
by the probability of detection ($\text{P}_d$)
and FAR.
The quantity~$\text{P}_d$
was obtained from the ratio between the number
of detected targets and the total numbers of known targets,
while FAR is defined by the number of false alarms
detected per square kilometer~\cite{Lundberg2006}.
Figure~\ref{f:roc} presents the
receiver operating
characteristic~(ROC)
curves~\cite{ROC}
of the change detection results, showing
the probability of detection versus the false alarm rates for the
different evaluated values of~$C$.
We compared the change detection results
obtained
from
the proposed method
with
the results
described
in~\cite{Ulander2005},~\cite{vu2017},
and~\cite{vu2018}.
The proposed method excels
in terms of
probability
of detection
and false alarm rate in comparison
to~\cite{Ulander2005,vu2017,vu2018}.

For example,
for a detection probability
of~$98\%$,
our proposed change detection method
presents~$\log_{10} (\text{FAR})$
about
$-0.5$,
while~\cite{Ulander2005,vu2017,vu2018}
have~$\log_{10} (\text{FAR})$
about~$1.4$,~$-0.3$,
and~$0.14$,
respectively.
For~$\log_{10} (\text{FAR}) = - 0.9$,
i.e., a very low FAR,
the probability detection
given by~\cite{Ulander2005}
drops
to~$60\%$,
while our proposal still maintains
the probability of detection more than~$90\%$.
The detection probability
of
our proposed method
and~\cite{vu2017}
reach~$100\%$
with~$\log_{10} (\text{FAR}) \approx 1$,
while~\cite{Ulander2005}
and~\cite{vu2018}
have full detection
for~$\log_{10} (\text{FAR}) \approx 1.5$
and~$\log_{10} (\text{FAR}) \approx 2$,
respectively.
Additionally,
detection probability improvements
of our method compared to~\cite{vu2017}
are found in the range of~$(0.93,0.98)$.
For example,
for a probability of detection of~$0.97\%$,
our proposed change detection method
presents~$\log_{10} (\text{FAR})$
about
$-0.8$,
while~\cite{vu2017}
has~$\log_{10} (\text{FAR}) \approx -0.2$.

\section{Conclusion}
\label{s:concl}

In this paper,
we presented five methods
to obtain ground
scene prediction of SAR images
based on image stack.
The experimental results revealed
that among the
considered
techniques,
the median method
yielded
the most accurate
ground
prediction.
The statistical
characteristics
of the obtained~GSP image
were
similar to the image of interest.
Moreover,
the median method
excels
in terms of
quality adjustment measures,
and the changes
in the image stack
were not visually presented
in the predicted image.
The
GSP image
based on the method  was used as a reference image in
a CDA,
presenting competitive
performance when compared
with recently published results.

\section*{Acknowledgments}
This study was financed in part by the
Conselho Nacional de Desenvolvimento Cient\'ifico e Tecnol\'ogico,
CNPq, Brazil,
Coordena\c{c}\~ao de Aperfei\c{c}oamento de Pessoal de N\'ivel Superior,
CAPES, Brazil,
Swedish-Brazilian Research and Innovation Centre (CISB),
and Saab AB.

{\small
\singlespacing
\bibliographystyle{siam}
\bibliography{bib}

\begin{thebibliography}{10}

\bibitem{ahmed2013removal}
{\sc F.~Ahmed and S.~Das}, {\em Removal of high-density salt-and-pepper noise
  in images with an iterative adaptive fuzzy filter using alpha-trimmed mean},
  IEEE Transactions on Fuzzy Systems, 22 (2013), pp.~1352--1358.

\bibitem{dimas2019}
{\sc D.~I. Alves, B.~G. Palm, M.~I. Pettersson, V.~T. Vu, R.~Machado, B.~F.
  Uchoa-Filho, P.~Dammert, and H.~Hellsten}, {\em A statistical analysis for
  wavelength-resolution {SAR} image stacks}, IEEE Geoscience and Remote Sensing
  Letters, 17 (2019), pp.~227--231.

\bibitem{amirmazlaghani2009}
{\sc M.~Amirmazlaghani, H.~Amindavar, and A.~Moghaddamjoo}, {\em Speckle
  suppression in {SAR} images using the 2-{D} {GARCH} model}, IEEE Transactions
  on Image Processing, 18 (2009), pp.~250--259.

\bibitem{baselice2013markovian}
{\sc F.~Baselice, G.~Ferraioli, and V.~Pascazio}, {\em Markovian change
  detection of urban areas using very high resolution complex {SAR} images},
  IEEE Geoscience and Remote Sensing Letters, 11 (2013), pp.~995--999.

\bibitem{belcher2013}
{\sc D.~P. Belcher and P.~S. Cannon}, {\em Ionospheric effects on synthetic
  aperture radar ({SAR}) clutter statistics}, IET Radar, Sonar \& Navigation, 7
  (2013), pp.~1004--1011.

\bibitem{biscainho2004}
{\sc L.~W. Biscainho}, {\em {AR} model estimation from quantized signals}, IEEE
  Signal Processing Letters, 11 (2004), pp.~183--185.

\bibitem{Brockwell2013}
{\sc P.~J. Brockwell and R.~A. Davis}, {\em Time series: {T}heory and methods},
  Springer, 2013.

\bibitem{Brockwell2016}
\leavevmode\vrule height 2pt depth -1.6pt width 23pt, {\em Introduction to time
  series and forecasting}, Springer, 2016.

\bibitem{bustos2009}
{\sc O.~Bustos, S.~Ojeda, R.~Vallejos, et~al.}, {\em Spatial {ARMA} models and
  its applications to image filtering}, Brazilian Journal of Probability and
  Statistics, 23 (2009), pp.~141--165.

\bibitem{chen2013blind}
{\sc C.~Chen, J.~Ni, and J.~Huang}, {\em Blind detection of median filtering in
  digital images: {A} difference domain based approach}, IEEE Transactions on
  Image Processing, 22 (2013), pp.~4699--4710.

\bibitem{chen2015median}
{\sc J.~Chen, X.~Kang, Y.~Liu, and Z.~J. Wang}, {\em Median filtering forensics
  based on convolutional neural networks}, IEEE Signal Processing Letters, 22
  (2015), pp.~1849--1853.

\bibitem{Cintra2013}
{\sc R.~J. Cintra, A.~C. Frery, and A.~D. Nascimento}, {\em Parametric and
  nonparametric tests for speckled imagery}, Pattern Analysis and Applications,
  16 (2013), pp.~141--161.

\bibitem{eltoft1998}
{\sc T.~Eltoft and K.~A. Hogda}, {\em Non-gaussian signal statistics in ocean
  {SAR} imagery}, IEEE Transactions on Geoscience and Remote Sensing, 36
  (1998), pp.~562--575.

\bibitem{ghirmai2015}
{\sc T.~Ghirmai}, {\em Representing a cascade of complex {G}aussian {AR} models
  by a single {L}aplace {AR} model}, IEEE Signal Processing Letters, 22 (2015),
  pp.~110--114.

\bibitem{gonzalez2008}
{\sc R.~C. Gonzalez and R.~Woods}, {\em Digital image processing}, 2008.

\bibitem{gudnason2009}
{\sc J.~Gudnason, J.~Cui, and M.~Brookes}, {\em {HRR} automatic target
  recognition from superresolution scattering center features}, IEEE
  Transactions on Aerospace and Electronic Systems, 45 (2009).

\bibitem{hampel2011}
{\sc F.~R. Hampel, E.~M. Ronchetti, P.~J. Rousseeuw, and W.~A. Stahel}, {\em
  Robust statistics: {T}he approach based on influence functions}, Wiley Online
  Library, 2011.

\bibitem{hellsten1996}
{\sc H.~Hellsten, L.~M. Ulander, A.~Gustavsson, and B.~Larsson}, {\em
  Development of {VHF} {CARABAS} {II} {SAR}}, in Radar Sensor Technology,
  vol.~2747, 1996, pp.~48--61.

\bibitem{hoekman2000l}
{\sc D.~H. Hoekman and M.~J. Quiriones}, {\em Land cover type and biomass
  classification using {A}ir{SAR} data for evaluation of monitoring scenarios
  in the {C}olombian {A}mazon}, IEEE Transactions on Geoscience and Remote
  Sensing, 38 (2000), pp.~685--696.

\bibitem{hyndman2006}
{\sc R.~J. Hyndman and A.~B. Koehler}, {\em Another look at measures of
  forecast accuracy}, International Journal of Forecasting, 22 (2006),
  pp.~679--688.

\bibitem{inglada2007}
{\sc J.~Inglada and G.~Mercier}, {\em A new statistical similarity measure for
  change detection in multitemporal {SAR} images and its extension to
  multiscale change analysis}, IEEE Transactions on Geoscience and Remote
  Sensing, 45 (2007), pp.~1432--1445.

\bibitem{Kay1998-2}
{\sc S.~M. Kay}, {\em Fundamentals of statistical signal processing:
  {D}etection theory}, vol.~II, Prentice Hall, 1998.

\bibitem{kirchner2010detection}
{\sc M.~Kirchner and J.~Fridrich}, {\em On detection of median filtering in
  digital images}, in Media forensics and security II, vol.~7541, International
  Society for Optics and Photonics, 2010, p.~754110.

\bibitem{liu2014}
{\sc B.~Liu, V.~G. Reju, and A.~W. Khong}, {\em A linear source recovery method
  for underdetermined mixtures of uncorrelated {AR}-model signals without
  sparseness}, IEEE Transactions on Signal Processing, 62 (2014),
  pp.~4947--4958.

\bibitem{Lundberg2006}
{\sc M.~Lundberg, L.~M. Ulander, W.~E. Pierson, and A.~Gustavsson}, {\em A
  challenge problem for detection of targets in foliage}, in
  Proceedings-Algorithms for Synthetic Aperture Radar Imagery XIII, vol.~6237,
  2006.

\bibitem{Renato2016}
{\sc R.~Machado, V.~T. Vu, M.~I. Pettersson, P.~Dammert, and H.~Hellsten}, {\em
  The stability of {UWB} low-frequency {SAR} images}, IEEE Geoscience and
  Remote Sensing Letters, 13 (2016), pp.~1114--1118.

\bibitem{robust}
{\sc R.~A. Maronna, R.~D. Martin, V.~J. Yohai, and M.~Salibi{\'a}n-Barrera},
  {\em Robust statistics: {T}heory and methods (with {R})}, Wiley, 2018.

\bibitem{mercier2008}
{\sc G.~Mercier, G.~Moser, and S.~B. Serpico}, {\em Conditional copulas for
  change detection in heterogeneous remote sensing images}, IEEE Transactions
  on Geoscience and Remote Sensing, 46 (2008), pp.~1428--1441.

\bibitem{ROC}
{\sc C.~E. Metz}, {\em Basic principles of {ROC} analysis}, in Seminars in
  nuclear medicine, vol.~8, Elsevier, 1978, pp.~283--298.

\bibitem{milenkovic1986}
{\sc P.~Milenkovic}, {\em Glottal inverse filtering by joint estimation of an
  {AR} system with a linear input model}, IEEE transactions on acoustics,
  speech, and signal processing, 34 (1986), pp.~28--42.

\bibitem{montazeri2016three}
{\sc S.~Montazeri, X.~X. Zhu, M.~Eineder, and R.~Bamler}, {\em
  Three-dimensional deformation monitoring of urban infrastructure by
  tomographic {SAR} using multitrack {T}erra{SAR-X} data stacks}, IEEE
  Transactions on Geoscience and Remote Sensing, 54 (2016), pp.~6868--6878.

\bibitem{oten2004adaptive}
{\sc R.~Oten and R.~J. de~Figueiredo}, {\em Adaptive alpha-trimmed mean filters
  under deviations from assumed noise model}, IEEE Transactions on Image
  Processing, 13 (2004), pp.~627--639.

\bibitem{Palm2018}
{\sc B.~G. Palm, D.~I. Alves, V.~T. Vu, M.~I. Pettersson, F.~M. Bayer, R.~J.
  Cintra, R.~Machado, P.~Dammert, and H.~Hellsten}, {\em Autoregressive model
  for multi-pass {SAR} change detection based on image stacks}, in
  Proceedings-Image and Signal Processing for Remote Sensing XXIV, vol.~10789,
  International Society for Optics and Photonics, 2018.

\bibitem{Palm2019}
{\sc B.~G. Palm, F.~M. Bayer, R.~J. Cintra, M.~I. Pettersson, and R.~Machado},
  {\em Rayleigh regression model for ground type detection in {SAR} imagery},
  IEEE Geoscience and Remote Sensing Letters, 16 (2019), pp.~1660 -- 1664.

\bibitem{data}
{\sc SDMS}, {\em {S}ensor {D}ata {M}anagement {S}ystem public web site},
  (2018).
\newblock https://www.sdms.afrl.af.mil/index.php.

\bibitem{sportouche2017}
{\sc H.~Sportouche, J.-M. Nicolas, and F.~Tupin}, {\em Mimic capacity of fisher
  and generalized gamma distributions for high-resolution {SAR} image
  statistical modeling}, IEEE Journal of Selected Topics in Applied Earth
  Observations and Remote Sensing, 10 (2017), pp.~5695--5711.

\bibitem{tison2004}
{\sc C.~Tison, J.-M. Nicolas, F.~Tupin, and H.~Ma{\^\i}tre}, {\em A new
  statistical model for {M}arkovian classification of urban areas in
  high-resolution {SAR} images}, IEEE Transactions on Geoscience and Remote
  Sensing, 42 (2004), pp.~2046--2057.

\bibitem{ulander2001}
{\sc L.~Ulander, P.-O. Frolind, A.~Gustavsson, H.~Hellsten, T.~Jonsson,
  B.~Larsson, and G.~Stenstrom}, {\em Performance of the {CARABAS-II}
  {VHF}-band synthetic aperture radar}, in Geoscience and Remote Sensing
  Symposium, 2001. IGARSS'01. IEEE 2001 International, vol.~1, IEEE, 2001,
  pp.~129--131.

\bibitem{Ulander2005}
{\sc L.~M. Ulander, M.~Lundberg, W.~Pierson, and A.~Gustavsson}, {\em Change
  detection for low-frequency {SAR} ground surveillance}, IEEE
  Proceedings-Radar, Sonar and Navigation, 152 (2005), pp.~413--420.

\bibitem{ulander2004}
{\sc L.~M. Ulander, W.~E. Pierson, M.~Lundberg, P.~Follo, P.-O. Frolind, and
  A.~Gustavsson}, {\em Performance of {VHF}-band {SAR} change detection for
  wide-area surveillance of concealed ground targets}, in Algorithms for
  Synthetic Aperture Radar Imagery XI, vol.~5427, International Society for
  Optics and Photonics, 2004, pp.~259--271.

\bibitem{vu2017}
{\sc V.~T. Vu}, {\em Wavelength-resolution {SAR} incoherent change detection
  based on image stack}, IEEE Geoscience and Remote Sensing Letters, 14 (2017),
  pp.~1012--1016.

\bibitem{vu2018}
{\sc V.~T. Vu, N.~R. Gomes, M.~I. Pettersson, P.~Dammert, and H.~Hellsten},
  {\em Bivariate gamma distribution for wavelength-resolution {SAR} change
  detection}, IEEE Transactions on Geoscience and Remote Sensing,  (2018),
  pp.~1--9.

\bibitem{Viet2017}
{\sc V.~T. Vu, M.~I. Pettersson, R.~Machado, P.~Dammert, and H.~Hellsten}, {\em
  False alarm reduction in wavelength-resolution {SAR} change detection using
  adaptive noise canceler}, IEEE Transactions on Geoscience and Remote Sensing,
  55 (2017), pp.~591--599.

\bibitem{wang2014efficient}
{\sc Y.~Wang, X.~X. Zhu, and R.~Bamler}, {\em An efficient tomographic
  inversion approach for urban mapping using meter resolution {SAR} image
  stacks}, IEEE Geoscience and Remote Sensing Letters, 11 (2014),
  pp.~1250--1254.

\bibitem{wang1999progressive}
{\sc Z.~Wang and D.~Zhang}, {\em Progressive switching median filter for the
  removal of impulse noise from highly corrupted images}, IEEE Transactions on
  Circuits and Systems II: Analog and Digital Signal Processing, 46 (1999),
  pp.~78--80.

\bibitem{white1991}
{\sc R.~G. White}, {\em Change detection in sar imagery}, International Journal
  of remote sensing, 12 (1991), pp.~339--360.

\bibitem{zhang2014revealing}
{\sc Y.~Zhang, S.~Li, S.~Wang, and Y.~Q. Shi}, {\em Revealing the traces of
  median filtering using high-order local ternary patterns}, IEEE Signal
  Processing Letters, 21 (2014), pp.~275--279.

\bibitem{zheng2014}
{\sc Y.~Zheng, X.~Zhang, B.~Hou, and G.~Liu}, {\em Using combined difference
  image and $k$-means clustering for {SAR} image change detection}, IEEE
  Geoscience and Remote Sensing Letters, 11 (2014), pp.~691--695.

\end{thebibliography}
}

\end{document}